\documentclass[twocolumn,prl,10pt,aps,floatfix]{revtex4-1}
\usepackage{graphicx}
\usepackage{amsmath}
\usepackage{subcaption}
\usepackage[colorlinks,linkcolor=blue,citecolor=blue,urlcolor=blue]{hyperref}
\usepackage{color}

\begin{document}

\title{Operator growth 
in a quantum compass model\\ on a Bethe lattice}
\author{X. Zotos$^{1,2,3}$}
\affiliation{$^1$Department of Physics, 
University of Crete, 70013 Heraklion, Greece}
\affiliation{$^2$Foundation for Research and Technology - Hellas, 71110 
Heraklion, Greece}
\affiliation{
$^3$Leibniz Institute for Solid State and Materials Research IFW Dresden, 
01171 Dresden, Germany} 

\date{\today}

\begin{abstract}
The time evolution of local operators in quantum compass models 
is characterized by simplicity as it can be represented as 
expanding and contracting strings of operators. 
Here we present an analytical solution to the problem of  growth of 
a local energy operator in a quantum compass model on a Bethe lattice. 
We find a linear increase in time of the average operator length
and a diffusive spreading of the operator length distribution.
By a moment method we evaluate the local energy autocorrelation 
function that shows a Lorentzian shape at low frequencies.  
Furthermore, by a stochastic method we visualize the expansion of the 
string cloud.
\end{abstract}
\maketitle


The Bethe lattice \cite{bethe}, due to its distinctive topological structure, 
offers exact solutions to statistical mechanics problems. 
Over the last years, in studies on quantum 
chaos, there are very interesting propositions linking 
the growth of local operators in quantum many-body systems 
under unitary dynamics to the emergence of irreversibility and dissipative 
behavior. 
Diagnostics as the OTOC (out-of-time-order correlator) \cite{otoc}
have been extensively studied in a large variety of prototype models
\cite{syk, pollmann} in search of universal features.
The main prediction of these studies is that the 
operator evolution has a light-cone structure in space-time, while the front
broadens diffusively as a function of time. 
Furthermore, universal properties of operator growth  
have beed proposed \cite{altman} and a relation of the OTOC to the 
Loschmidt echo \cite{zurek}. 

From another perspective, the dynamics of quantum compass models 
is an old subject motivated by novel materials with intertwined 
spin and orbital degrees of freedom \cite{komskii}.
These, often two dimensional, quantum magnets are characterized by 
strongly anisotropic interactions \cite{brink} and 
they have recently been brought back into attention
as prototype, fictitious spin models for quantum computing \cite{kitaev}.

In this work, we study the time evolution of a local energy operator 
in a quantum compass model on a Bethe lattice. 
The key idea of our study is that the operators 
generated by the time evolution of a local energy operator have a very simple 
structure, as strings growing and contracting on the lattice, 
a {\it discrete quantum branching}.  
This allows us to analytically evaluate the average size of strings and 
their distribution as a function of time and by a moment method 
the energy autocorrelation function at infinite temperature.
Last but not least, by a stochastic approach we provide a picture 
of the expanding {\it string cloud}.

\begin{figure}[!h]
\includegraphics[width=1.0\linewidth, angle=0]{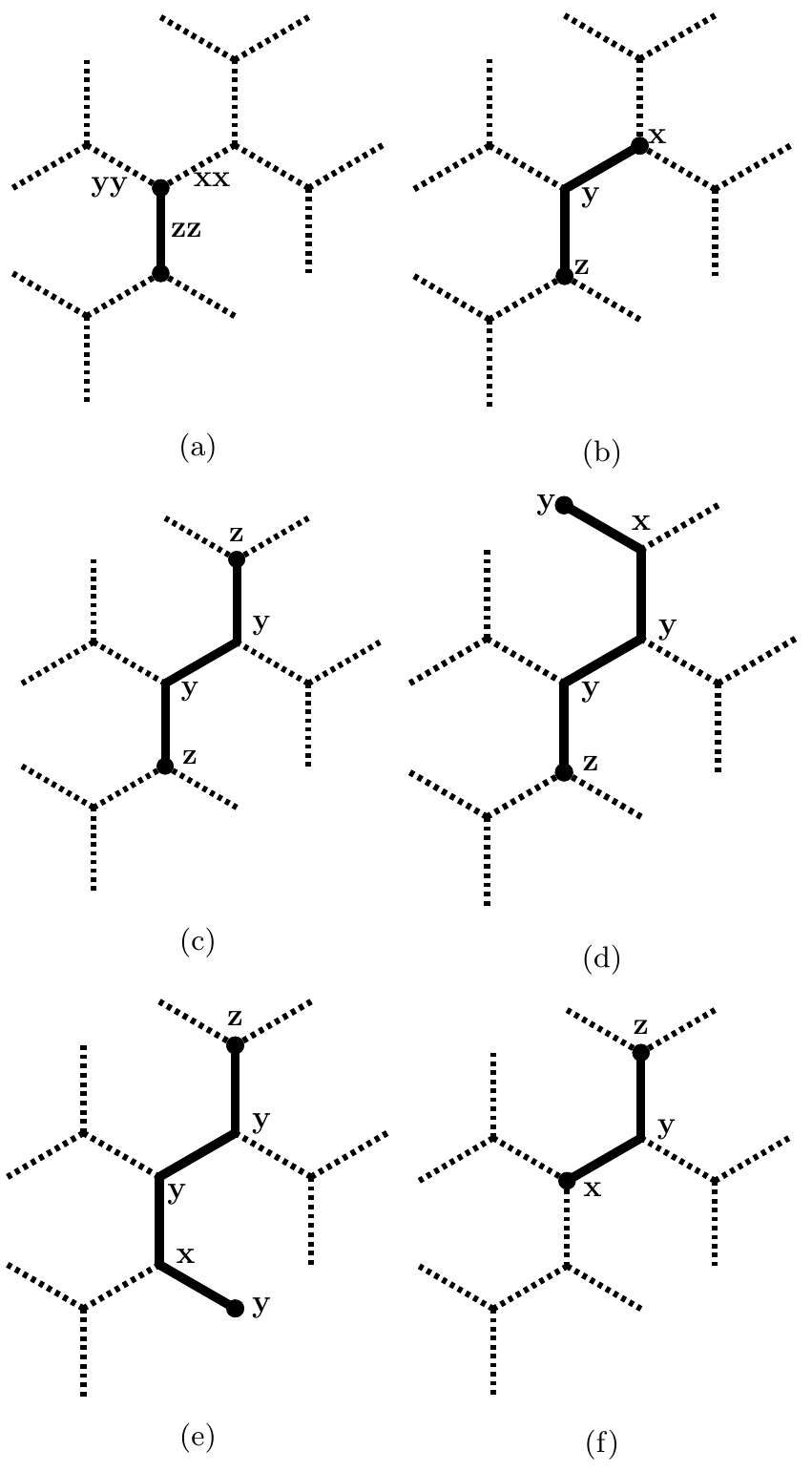}
\caption{Example of successive string configurations,\\
(a) $O=\tau_0^z\tau_1^z$,
(b) ${\cal L} O$, (c) ${\cal L}^2 O$, (d,e,f) ${\cal L}^3 O$. }
\label{fig1}
\end{figure}

\bigskip
We study the quantum compass  model on a Bethe lattice with threefold 
coordination depicted in Fig.\ref{fig1}a and given by the Hamiltonian,
\begin{eqnarray}
H&=&\sum_{<ij>_x} h_{ij}^x +\sum_{<ij>_y} h_{ij}^y
+\sum_{<ij>_z} h_{ij}^z
\nonumber\\
&=&-J\sum_{<ij>_x} \tau_i^x\tau_j^x
  -J\sum_{<ij>_y} \tau_i^y\tau_j^y
  -J\sum_{<ij>_z} \tau_i^z\tau_j^z
\nonumber\\
\end{eqnarray}
\noindent
where $\tau^{x,y,z}$ are Pauli pseudospin-1/2 operators and 
$J$ is the unit of energy. 
In Fig.\ref{fig1}a $zz$ represents an initial local energy operator
$O=\tau_0^z\tau_1^z$. The time evolution $e^{iHt}Oe^{-iHt}$ 
of this operator is obtained 
by successive application of the ''Liouvilian" operator ${\cal L}$,
${\cal L}O=[H,O],~{\cal L}^2O=[H,[H,O]]$ etc.
In Figs.\ref{fig1}b,c,d,e we show examples of different orders ${\cal L}^m$. 
For instance the operator string in Fig.\ref{fig1}d is 
$\tau_0^z\tau_1^y\tau_2^y\tau_3^x\tau_4^y$ (the numbering of sites is 
indicative).
The key point is the observation 
that there is no branching of the generated strings and 
when a term of $H$ acts at the middle of a string, 
it is annihilated \cite{briffa}. Essentialy, the Bethe lattice has the 
same  local bond structure as the honeycomb lattice. 
Thus the strings generated can be easily visualized and accounted for 
as strings expanding by one leg in 4 possible directions at their ends 
or contracting by one leg at either end. 
It is easy to assign the operator at each 
node as, by construction, it is the end of the corresponding bond. 
For instance, 
in the operator ${\cal L}^2 O$ there are 24 strings, 4 of length-1 
at the origin, 4 of length-1 in the nearest bonds and 16 of length-3.
Thus the evolution of the number $N_l^{m+1}$ of strings of length $l$ 
at the step ${\cal L}^{m+1}$ is given by the simple recursion relation 
\begin{equation}
N_l^{m+1}=2N_{l+1}^m+4N_{l-1}^m 
\label{recursion}
\end{equation}
with appropriate boundary condition for $N^m_1$, 
as the application of ${\cal L}$ also annihilates 
a length-1 string. It describes a discrete quantum branching.
Numerical iteration of (\ref{recursion}) generates 
$N^m_l$ to the desired order, but it is also easy (see Appendix) 
to obtain an analytical solution.

\begin{figure}[!h]
\begin{center}
\includegraphics[width=1\linewidth, angle=0]{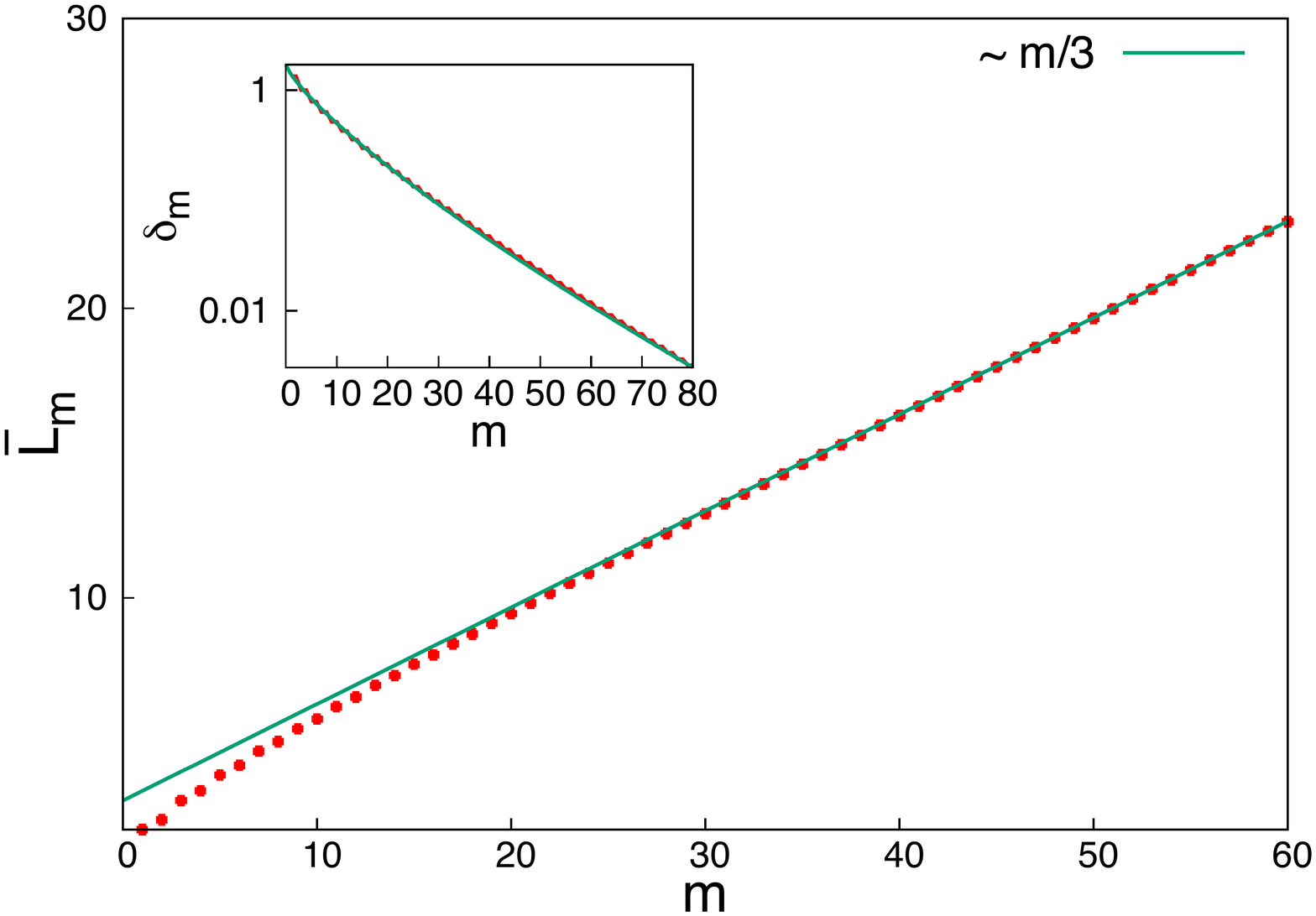}
\caption{Time dependence of average string length ${\bar L}_m$.
Inset: $\delta_m$ is the deviation from slope $m/3$, fitted 
to a stretched exponential.}
\label{fig2}
\end{center}
\end{figure}

\begin{figure}[!h]
\begin{center}
\includegraphics[width=1\linewidth, angle=0]{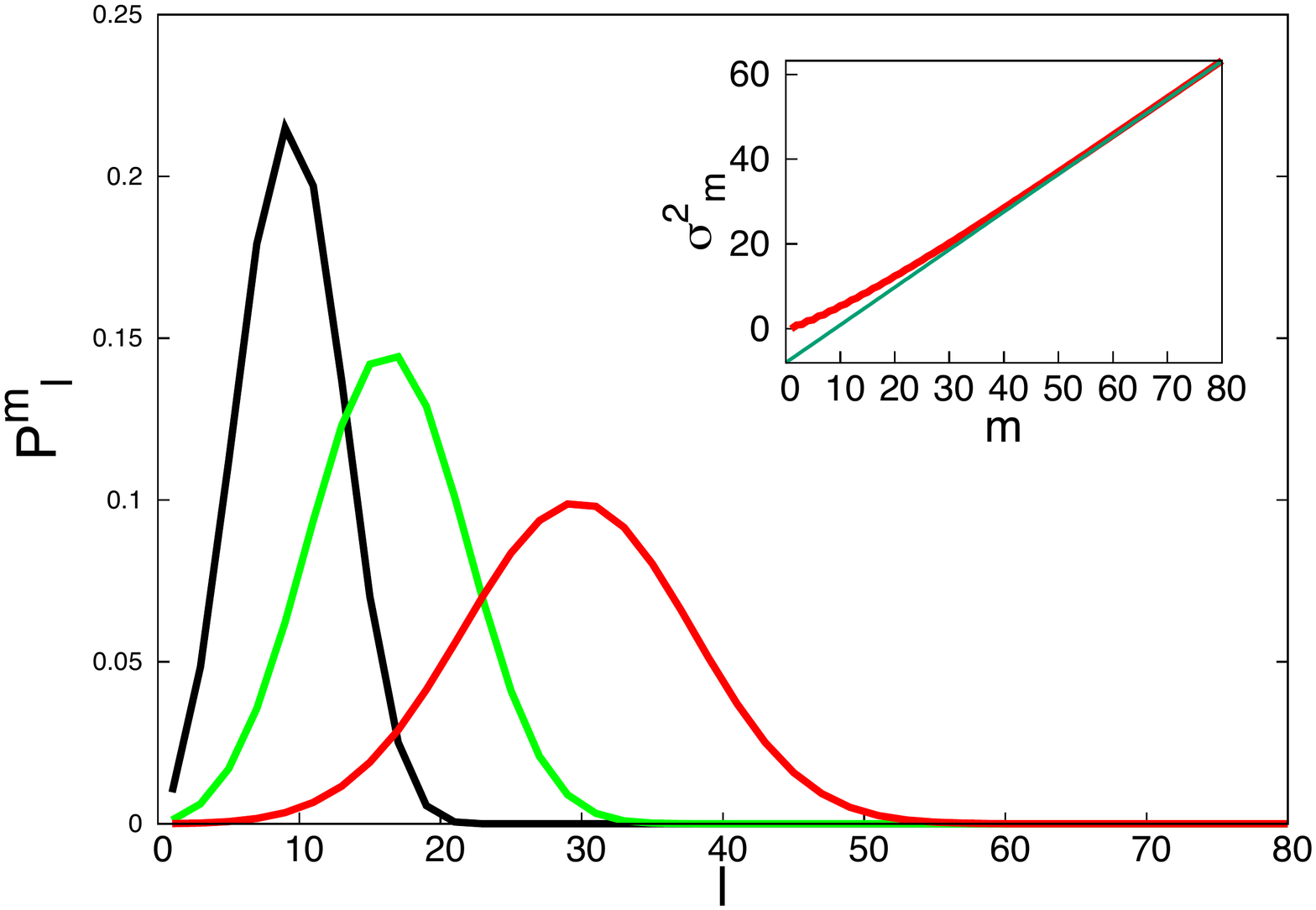}
\caption{String length distribution $P^m_l$, $m=20$ (black), $m=40$ 
(green), $m=80$ (red). Inset: width $\sigma^2_m$ of distribution $P^m_l$. }
\label{fig3}
\end{center}
\end{figure}

The essential features of the expanding string cloud is that, 
(i) after an initial transient of about $m\sim 20$, 
the average length $\bar L_m=\sum_l l P_l^m,~~~
(P^m_l= N_l^{m}/\sum_l N_l^{m})$ 
increases linearly as a function of Liouville time steps $m$ with slope
$1/3$, 
(ii) as shown in the inset of Fig.\ref{fig2}, 
the deviation $\delta_m={\bar L}_m-(m/3+{\rm constant})$ 
decreases as a stretched exponential $\sim e^{-0.2m^{0.788}}$,
(iii) the distribution of lengths $P^m_l$,
shown in Fig.\ref{fig3}, tends to a Gaussian with width 
$\sim\sqrt{m}$ for large $m$. The width of the distribution,
shown in the inset of Fig.\ref{fig3}, is given by
$\sigma_m^2=\sum_l(l-{\bar L}_m)^2 P^m_l$.

\begin{figure}[!h]
\begin{center}
\includegraphics[width=1\linewidth, angle=0]{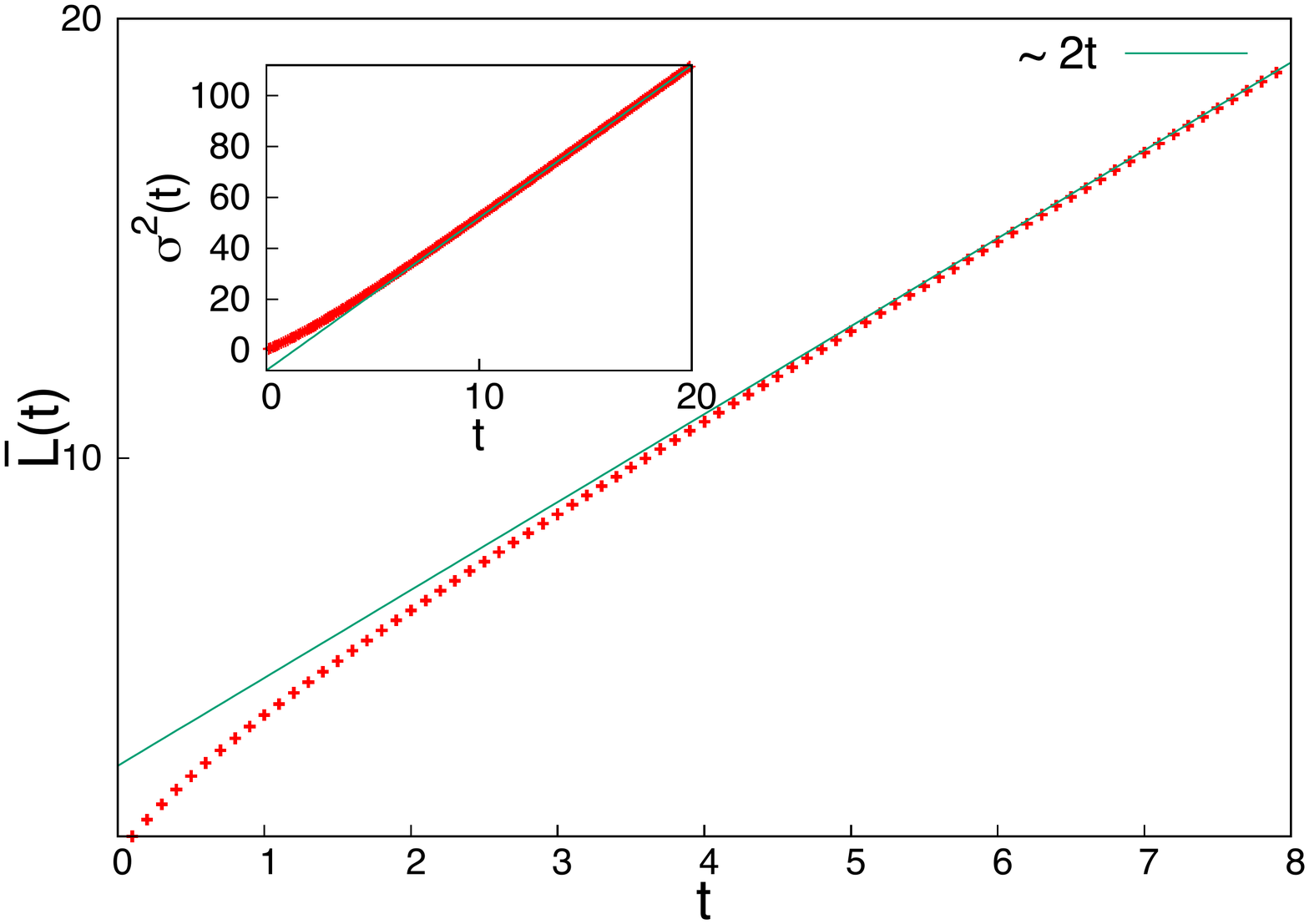}
\caption{Time dependence of average string length ${\bar L}(t)$
as a function of time.\\
Inset: the variance $\sigma^2(t)$ with slope $6$.}
\label{fig4}
\end{center}
\end{figure}

Furthermore, the average length as a function of time can be obtained 
from the moment expansion of the probability distribution,
\begin{equation}
p_l(t)=\sum_{m=0}^{+\infty} \frac{N_l^m}{m!}t^m.
\end{equation}
\noindent
The normalized distribution $P_l(t)=p_l(t)/\sum_l p_l(t)$ gives the average 
length ${\bar L}(t)=\sum_l l P_l(t)$. 
As shown in Fig.\ref{fig4},
after a transient, the average length of strings 
${\bar L}(t)$ grows linearly with time $t$ with slope 2.

We should note that, in the enumeration of generated strings we do not take 
into account the minus sign (due the Pauli pseudospin-1/2 
commutation relations) that appears when a string is added in the 
list of strings of given configuration but with "head" and "tail" reversed.
We have verified that taking into account these minus signs, appearing 
to higher iteration order, does not 
qualitatively change the above operator growth picture.

\bigskip
Next, to correlate the diffusive growth of operator length distribution
to energy transport, we study the frequency dependenct of the local 
energy autocorrelation 
function, 
\begin{equation}
C(t)=< O(t)O>
\end{equation}
\noindent
in the infinite temperature limit, on a lattice with $L$ spins,
\begin{equation}
C(t)=\frac{1}{2^L} {\textrm tr}~O(t)O,
\end{equation}
\noindent
from a moment expansion,
\begin{equation}
C(t)=\sum_{m=0}^{\infty} \frac{(-1)^m}{(2m)!} \mu_{2m} t^{2m},~~
\mu_{2m}=\frac{1}{2^L} {\textrm tr}~O {\cal L}^{2m} O.
\label{moments}
\end{equation}

\noindent
As $O^2=1$, 
the autocorrelation function reduces to the evaluation of the
moments $\mu_{2m}$\cite{muller}, related to the number ${\bar \mu}_{2m}$ 
of strings of length one at the origin after $m$ Liouville steps. 
For instance 
the second moment is equal to 4 as there are 4 ways to return to the 
original string at the origin after ${\cal L}^2 O$. As we cannot enumerate 
the moments in an analytical way, we apply a complete 
generation and accounting of strings generated on a finite Bethe 
lattice with up to 12 branching steps in each direction. 
In this calculation, we take into account the minus signs appearing in the 
generated operators through the process discussed above. Furthermore, 

\begin{equation}
S(\omega)=\int_{-\infty}^{+\infty} C(t) e^{+i\omega t} dt
\end{equation}

\noindent
is evaluated by an extension to complex frequencies $z$,
\begin{equation}
c(z)=\int_0^{+\infty} C(t) e^{-z t} dt,~~\Re(z) > 0,
\end{equation}
\begin{equation}
S(\omega)=\lim_{\eta \rightarrow 0^+} 2\Re [c(\eta -i \omega)] 
\end{equation}
\noindent
and then $c(z)$ is conveniently expressed as a continued fraction
expansion,
\begin{equation}
c(z)=\frac{1}{z+\frac{\Delta_1}{z+\frac{\Delta_2}{z+..}}}.
\end{equation}
\noindent
The coefficients $\Delta_n$ are related to the moments $\mu_{2m}$ by
recursion relations \cite{muller}. 
A list of the first 12 moments growing as ${\bar \mu}_m\sim e^{1.5m}$ 
and corresponding $\Delta$ coefficients 
is given in the Table
(note that $\mu_m=2^m{\bar \mu}_m$, the factor of $2^m$ coming 
from the commutation relation of the Pauli matrices).
\begin{center}
\begin{tabular}{cccccc}
${\bar \mu}_2$ &${\bar \mu}_4$ &${\bar \mu}_6$ & ${\bar \mu}_8$ 
&${\bar \mu}_{10}$ &${\bar \mu}_{12}$ \\
\hline
4     &44    &676   &12316 & 249044 & 5404780 \\
$\Delta_1$ &$\Delta_2$ &$\Delta_3$ &$\Delta_4$ &$\Delta_5$ &$\Delta_6$ \\
\hline
16.00     &28.00    &27.43   &30.24  & 29.67 & 30.96 \\
\hline
\end{tabular}
\end{center}
\noindent

\begin{figure}[!h]
\begin{center}
\includegraphics[width=1\linewidth, angle=0]{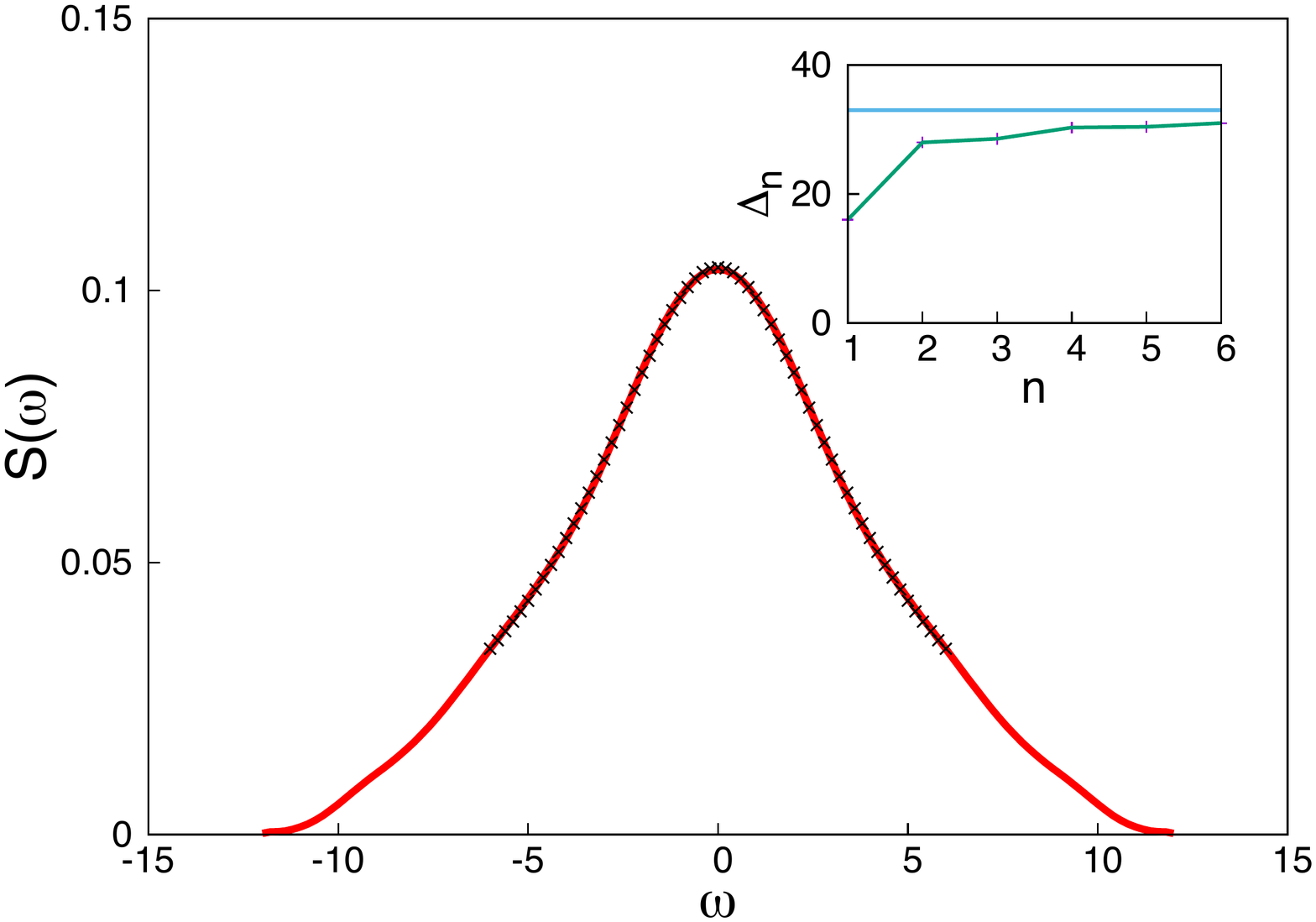}
\caption{Autocorrelation function $S(\omega)$ and a low frequency 
Lorentzian fit (black symbols).
Inset: coefficients of the continued fraction expansion and 
an asymptotic.}
\label{fig5}
\end{center}
\end{figure}

In Fig.\ref{fig5} we show $S(\omega)$ obtained from the continued fraction 
with a broadening $\eta=0.1$ and taking an asymptotic $\Delta_{n>6}=33$.
The low frequency behavior seems to be well fitted by a Lorentzian,  
further moments are necessary to confirm the diffusive 
character (altough the overall shape does not seem very sensitive 
to the exact value of the last moments and asymptotic value). 
It is an interesting issue whether  
the dissipative energy transport is related to 
the diffusive spreading of the string lengths distribution. 

As we are dealing with a growth problem, we can apply 
a stochastic approach to get a picture of the evolution of the string cloud.
Starting from the initial $zz$-bond, by a random choice of direction 
of expansion or contraction of a string, 
we stochastically generate a large sample of 
strings. Their number of a given configuration 
at each level of iteration is of course 
proportional to their number in the complete evolution of the string cloud. 
For instance, we can get an estimate of the moment 
${\bar \mu}_m\simeq N^m \cdot \frac{{\tilde N}^m}{N_s}$ from the known  
total number of strings $N^m$ at iteration $m$,
$N_s$ the number in the sample 
of randomly generated strings after $m$ iterations
and ${\tilde N}^m$ the number of $zz$-strings at the origin. 
This estimate, with an error $O(\frac{1}{\sqrt{N_s}})$, although 
can be very accurate, we found that 
it is not sufficient for the evaluation of the 
$\Delta_{n > 6}$'s in the continued fraction expansion, as 
the recursion relations are a highly unstable procedure.

To create the randomly generated strings, we first code the nodes of the Bethe 
lattice with a pair in coordinates $(i,j)$ where $i$ is the "radial" distance 
from the origin $(1,1)$ and $j$ the "circular" coordinate, e.g. 
$(1,1)$ is the origin, $(2,1),(2,2),(2,3)$ its nearest neighbors, 
$(3,1),(3,2)$ the neighbors of $(2,1)$, $(3,3),(3,4)$ the neighbors of 
$(2,2)$ etc.  Thus the number of nodes at level $i$ is equal 
to $3\cdot 2^{i-2}$. 
In the Bethe lattice, a string is uniquely defined by the coordinates 
of the nodes at its two ends $(i_b,j_b),(i_e,j_e)$.
So next, in each iteration we move stochastically one of the two ends 
in one of the three possible directions.

\begin{figure}[!h]
\includegraphics[width=1.0\linewidth, angle=0]{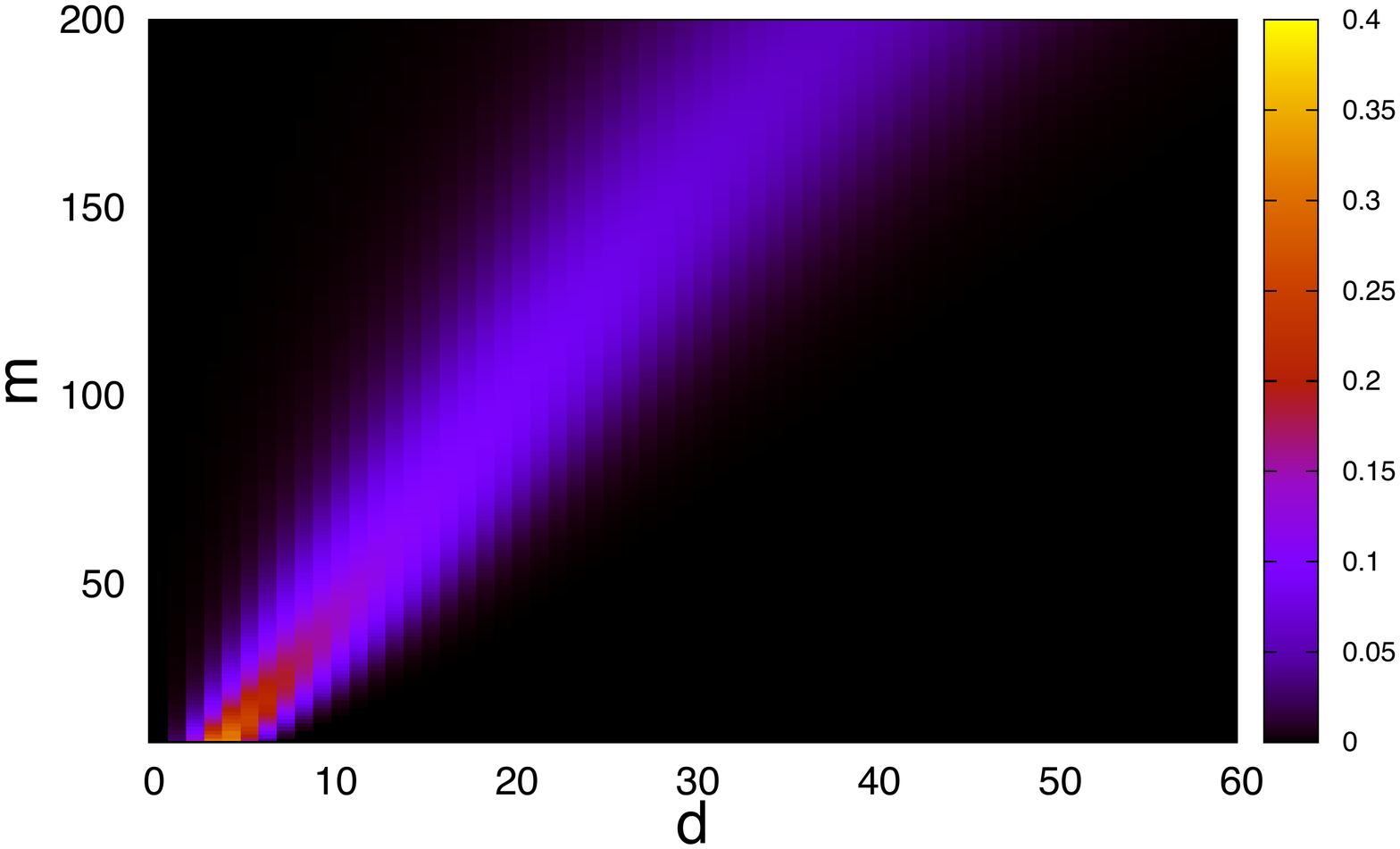}
\caption{Color map of the expansion of string cloud. 
The color intensity is proportional to the number of strings, 
normalized to the number of samples,
at average distance $d$ from the origin after $m$ (for clarity $m>4$) 
iterations. 
The distance $d$ in the
$x$-axis, the number of iterations $m$ in the $y$-axis.}
\label{fig6}
\end{figure}

To study the  evolution 
of the string cloud we can define an average distance of a string 
from the origin as $d=(i_b+i_e)/2$.
In Fig.\ref{fig6}, a picture of the expansion 
of the string cloud shows the same features 
as the above analysis of average length growth and distribution. 
We typically consider about $10^8$  
random string configurations. 

\bigskip
In conclusion, this quantum compass branching 
model is a paradigm of operator growth.
It demonstrates the linear in time evolution of 
the average length of operator strings, the diffusive spread of strings lengths
and the expansion of the string cloud. The simplicity of the string 
structure is due to the fully anisotropic compass interactions 
on the Bethe lattice.
Despite this apparent simplicity in the description 
of the string cloud, a complexity emerges in the distribution of 
string lengths and positions that leads to a diffusive growth of string 
lengths. It is an open question whether 
this complexity implies the diffusive  
transport we find in the energy autocorrelation function.
Addition of a magnetic field or "bond disorder", for instance replacing  a 
$zz$-bond by an $xx$-bond, only creates side-branching.  
In constrast, the addition of a different type of interaction, 
e.g. Heisenberg term, destroys the string structure.
By the stochastic approach as well as analytical methods, a further 
study of the operator growth as a statistical mechanics problem 
should be possible. 

\bigskip
\section{Acknowledgments}
This work was supported by the 
Deutsche Forschungsgemeinschaft through Grant HE3439/13.

\section{Appendix}
The recursion relation (\ref{recursion}) can be seen as the succesive 
application of 
a tridiagonal Toeplitz matrix $U$ of dimension $n$, 
with elements $a=4$ above the diagonal and  $b=2$ below,
on an initial vector $N^0=(1,0,0,0,....)^T$,
The right eigenvectors of $U$ are given by, 
$|x>_k=\frac{2}{n+1}\sqrt\frac{a}{b}\sin\frac{jk\pi}{n+1},~~j,k=1,...,n$
the left ones by, 
$<x|_k=\frac{2}{n+1}\sqrt\frac{b}{a}\sin\frac{jk\pi}{n+1},~~j,k=1,...,n$
and the corresponding eigenvalues $\epsilon_k=2\sqrt{ab}\cos\frac{k\pi}{n+1}$. 
Thus the string length vector $N^m_l,~l=1,...,m+1$ obtained from
$U^mN^0=N^m,~n > m+1$, has components 
\begin{eqnarray}
N^m_l&=&\sum_{k=1}^n \Big(2\sqrt{ab}\cos\frac{k\pi}{n+1}\Big)^m
\frac{2}{n+1}\Big(\frac{b}{a}\Big)^{1/2}\cdot
\nonumber\\
&&\sin\frac{k\pi}{n+1}\Big(\frac{a}{b}\Big)^{l/2}\sin\frac{kl\pi}{n+1}.
\end{eqnarray}
Taking $n\rightarrow \infty$ we obtain,
\begin{eqnarray}
N^m_l &=& \Big( \frac{2}{\pi} \Big) 
\int_{0}^{\pi} dx  \Big( 2\sqrt{ab}\cos x \Big)^m 
\Big( \frac{b}{a} \Big)^{1/2} \sin x \cdot 
\nonumber\\
&&\Big( \frac{a}{b} \Big)^{l/2} \sin(lx).
\label{nml}
\end{eqnarray}
\noindent
This expression is nonzero for 
$m~{\rm even}, l~{\rm odd}$ 
or  $m~{\rm odd}, l~ {\rm even}$ 
and $l \leq m+1$. Concretely, for $m~{\rm even}, l~{\rm odd}$ 
\begin{eqnarray}
N^m_l &=& \Big( \frac{2}{\pi} \Big) 
\Big( 2\sqrt{ab}\Big)^m 
\Big( \frac{b}{a} \Big)^{1/2} 
\Big( \frac{a}{b} \Big)^{l/2}\cdot 
\nonumber\\
&&\frac{l\pi}{2^{m+1}}
\frac{m!}
{ 
(\frac{m}{2}-\frac{l-1}{2})! 
(\frac{m}{2}+\frac{l-1}{2}+3)!
}. 
\label{nmlf}
\end{eqnarray}
\noindent
In the limit of large $m$ (\ref{nmlf}) 
can be evaluated in the saddle point approximation,
\begin{equation}
N^m_l\sim e^{-\frac{(l-m/3)^2}{2\sigma^2}},~~\sigma^2=\frac{8}{9}m.
\end{equation}

\end{document}